\documentclass[aps,preprint,prb,floatfix,showkeys]{revtex4}
\usepackage{graphicx}
\usepackage{multirow}
\usepackage{color}

\begin{document}

\title{The Structural and Electronic Properties of Pristine and Doped Polythiophene: Periodic Versus Molecular Calculations}

\author{Thaneshwor P. Kaloni}
%\email{thaneshwor.kaloni@kaust.edu.sa}
\author{Georg Schreckenbach}
\email{schrecke@cc.umanitoba.ca, +1-204-474-6261}
\author{Michael S. Freund}
\email{michael.freund@umanitoba.ca, +1-204-474-8772}
\affiliation{Department of Chemistry, University of Manitoba, Winnipeg, MB, R3T 2N2, Canada}

\begin{abstract}
Based on density functional theory calculations, the structural and electronic properties of polythiophene in periodic and oligomer forms have been investigated. In particular, the effects of Li or Cl adsorption onto a monolayer and Li or Cl-intercalation into bulk or bilayer polythiophene are addressed using periodic calculations. The binding energy of Li or Cl adsorbed bulk or bilayer polythiophene is significantly larger than for the monolayer. The trends in the binding energy as a function of adsorbent remain the same for both the periodic and molecular cases. The band gap or HOMO-LUMO gap and charge transfer are analysed. In addition, for the bulk or bilayer, different kinds of stacking have been considered. It is found that the parallel bulk or bilayer structure is energetically favorable compared to flipping the second layer by 180$^\circ$. This has been considered for both the periodic and oligomer forms. Moreover, for Li adsorption, polarons are found to be more stable than bipolarons, while the situation is opposite for Cl adsorption. The detailed analysis of the present study will be useful for understanding the structural properties and the tuneability of the electronic states, which is an important step to construct polythiophene based electronic devices.     

\end{abstract}
\maketitle
\section{Introduction}

Organic conjugated polymers have been studied widely in both experiment and theoretical modelling. Organic polymers are of practical interest due to the fact that they can potentially be used in electronic optical semiconductor devices \cite{Chiang,McGehee} as well as in sensors. \cite{ms2,s1,s2,s3,ms1} The structural and electronic properties of these polymers can be tuned and controlled easily by chemical modifications, such as doping.\cite{soth,Heeger} It has been reported that polyacetylene can be doped at room temperature with various dopants, which can tune the structural and electronic properties of conducting polymers.\cite{Chiang,Chiang1} The doping technique and mechanism have been widely used for a large number of conjugated polymers, and dopants play an important role in changing their properties from those of conventional covalent semiconductors.\cite{Chance-1986,ms-jacs} The doping can be achieved either by charge transfer or by application of an external electric field. By applying these doping techniques, the electronic and optical properties of conducting polymers can be engineered widely, including the transition from semiconductor to metal or further to insulator depending upon the dopant concentration.\cite{Heeger,Heeger2,Heeger3,moon,moon1} This induces an outstanding opportunity in switching between conducting and insulating properties, which indeed paves the way for applications in optoelectronics, such as organic polymer based transistors, photoresistances, light-emitting diodes, and thiophene based organic solar cells.\cite{Stevens}

It is well known that polyacetylene has a degenerate ground state with soliton formation. However, polythiophene is a general example of a polymer with a nondegenerate ground state. The charge transfer due to doping of the donors or acceptors produces polarons or bipolarons (depending on the dopant concentration) rather than soliton formation. Polythiophene and its derivatives are particularly useful for device applications.\cite{su,su1,jcp} Thus, it is important to understand the effects of donor and acceptor doping on polythiophene. Direct control over the band gap is highly desirable; it is an important issue to control or engineer the electronic structure of the polythiophene polymer. There have been several efforts to address the electronic structure of thiophenes using first-principles calculations.\cite{Salzner,Radhakrishnan,Radhakrishnan1,Kasai,Zade,Zamoshchik,Patra,Rittmeyer} In Ref.\ \cite{Rittmeyer}, modifications of the electronic properties of polythiophenes have been investigated by substituting a H atom with CH$_3$, NH$_2$, NO$_2$, and Cl in the framework of density functional theory (DFT). In particular, the influence of the above-mentioned substituents on the band gap has been investigated. Experimentally and theoretically, the optical properties of fully conjugated cyclo[n]thiophenes have been studied. Based on the absorption spectra, it has been demonstrated that the band gap and optical properties are closely related to the geometry of the thiophene molecule.\cite{Zhang,grey1} Thiophene-based materials are expected to be promising candidates in electronic memory devices, therefore, they have been the focus of instigation for the last two decades or so.\cite{Stevens,Rittmeyer} In addition, in case of conjugated $\pi$ systems, an annihilation of the electrons from the highest occupied molecular orbital (HOMO) and excitation of the electrons to the lowest unoccupied molecular orbital (LUMO) are achieved without difficulties.\cite{Reynolds}

Experimentally, dynamic doping of conjugated polymers can be achieved by creating a composite, which essentially contains a sufficient level of of immobilized anions and mobile cations to dope the polymer. By creating an interface between the doped conjugated polymer with a semiconductor that can be doped in the solid state, a field-driven change in conductivity can be achieved. Moreover, the electrodeposition of the heterojunctions can also be scalable upto nano-meter, which is enable to create devices on existing crossbar structures. \cite{ms4,Barman} Recently, using a conjugated polymer, a well-defined barrier to the ion drift has been demonstrated experimentally, which is expected to provide a mechanism for constructing a memory device with high performance.\cite{freund} It has been shown that the conjugated polymer barrier layer remains conducting at the interface between the metal oxide and conjugated polymer doped by Li$^+$ ion. This can be achieved by electro-deposition of polythiophene in the presence of dodecyl sulfate followed by a thin film of polypyrrole in the presence of dodecyl benzene sulfonate, where the electrochemical deposition of a thin film made of WO$_3$ is used to make a contact to the polypyrrole junction. However, the scaling behavior and the role of of the Li or Cl on the structural and electronic properties of polythiophene in various stacking geometries either in periodic or oligomer forms have not been studied so far. Therefore, in this paper, in the framework of first-principles density functional theory based model calculations, the structural and electronic properties of the periodic pristine monolayer polythiophene, bilayer polythiophene with parallel stacking and flipped by 180$^\circ$, and Li and Cl adsorption/intercalation in each case are investigated. These systems are also studied in molecular form. The obtained result demonstrate that the electronic structure can easily be tuned and controlled either by adsorption/intercalation or by creating multilayers in bulk or bilayer stacking sequences.

\section{Computational details}
All the periodic calculations were performed using density functional theory (DFT) within the generalized gradient approximation in the Perdew, Burke, and Ernzerhof parametrization \cite{pbe} as implemented in the QUANTUM-ESPRESSO package.\cite{paolo} The van der Waals interactions (DFT-D) are included in order to achieve an accurate description of the dispersion.\cite{grime} A relatively high plane wave cutoff energy of 952 eV and a Monkhorst-Pack $24\times1\times1$ k-mesh are employed. The considered supercell of polythiophene has lattice constants of $a=15.59$ \AA\ and $b=15.00$ \AA. A vacuum layer of 20 \AA\ is used in order to avoid artificial interactions due to the periodic images. The atomic positions are optimized until all the forces have converged to 0.0001 eV/\AA. In periodic calculations, three different systems are considered (i) monolayer; a single layer of polythiophene with a vacuum along $y$- and $z$-directions, (ii) bulk; periodic repetition of polythiophene along $x$- and $z$-directions, and (iii) bilayer; two layers of polythiophene with a vacuum along $y$- and $z$-directions.

The structures of molecular polythiophene (oligomers) were optimized using DFT-D (PBE) \cite{pbe} as implemented in the Amsterdam Density Functional package ADF.\cite{adf,adf1,adf2} All the calculations were performed using double zeta Slater-type orbital basis sets (DZ). Geometries were optimized until the energy was converged to $1\cdot 10^{-4}$ eV. The default integration accuracy parameter of 4.0 is used for geometry optimizations and single point calculations for all cases under study except for the Cl-intercalated bilayer systems. In case of Cl-intercalated bilayer systems, a numerical integration of 7.0 is used in order to achieve convergence.

\section{Periodic calculations}
\subsection{Structural analysis}

\begin{table}[h]
\begin{tabular}{|c|c|c|c|c|c|c|}
\hline
Parameter& \multicolumn{3}{|c|}{\multirow{1}{*}{Pristine}}&\multicolumn{2}{|c|}{\multirow{1}{*}{Li-adsorbing}}& Cl-adsorbing\\
\cline{2-7}
\hline
&\begin{tabular}{c} Present \\ work \end{tabular}&\begin{tabular}{c} Previous \\ report  \cite{jcp,Lischka} \end{tabular}&Expt.\ \cite{Casado}&\begin{tabular}{c} Present \\ work \end{tabular}&\begin{tabular}{c} Previous \\ report  \cite{jcp,Lischka} \end{tabular}&\begin{tabular}{c} Present \\ work \end{tabular} \\
\hline
C1C2&1.438&1.448&1.433&1.408&--&1.422 \\
\hline
C2C3&1.361&1.385&1.357 &1.418&1.424&1.401   \\
\hline
C3C4&1.410&1.416&--&1.395&1.391&1.393\\
\hline
C4C5&1.369&1.376&--&1.397&--&1.399\\
\hline
C5S1&1.713&1.720&1.717&1.752&--&1.793\\
\hline
C2S1&1.731&1.735&--&1.769&1.778&1.742\\
\hline
C1C8&1.361&1.352&--&1.410&1.403&1.399\\
\hline
C8C7&1.411&1.427&--&1.427&1.431&1.389\\
\hline
C7C6&1.363&1.353&--&1.411&1.410&1.398\\
\hline
C6S2&1.729&1.737&--&1.787&1.788&1.739\\
\hline
S2C1&1.731&1.737&--&1.786&1.789&1.738\\
\hline
S2Li/Cl&--&--&--&2.376&2.369&3.675\\
\hline
C6Li/Cl&--&--&--&2.235&2.245&3.490\\
\hline
C1Li/Cl&--&--&--&2.226&2.211&3.402\\
\hline
C7Li/Cl&--&--&--&2.175&2.190&3.268\\
\hline
C8Li/Cl&--&--&--&2.171&2.178&3.210\\
\hline
C5S1C2&91.50&90.80&91.90&92.35&89.90&92.71\\
\hline
C5C4C3&113.01&113.02&--&114.16&--&113.79\\
\hline
C4C3C2&113.01&113.41&--&113.61&--&113.71\\
\hline
C3C2C1&128.96&--&--&129.28&--&128.33\\
\hline
C1C8C7&113.99&--&--&114.39&--&113.64\\
\hline
C1S2C6&92.02&--&--&92.16&--&92.50\\
\hline
\end{tabular}
\caption{Calculated bond lengths (in \AA) and bond angles (in degree), periodic calculations} 
\end{table}

\begin{figure}[ht]
\includegraphics[width=0.6\textwidth,clip]{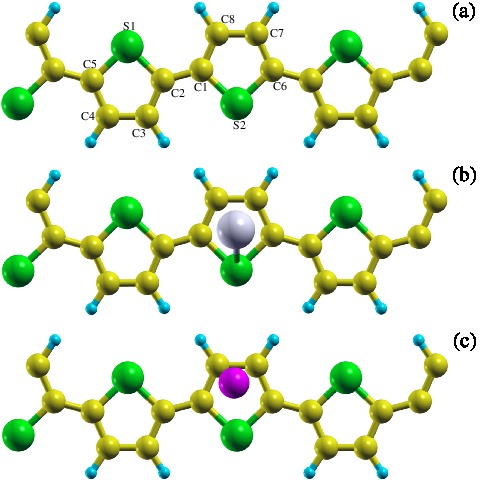}
\caption{The optimized structures of periodic polythiophene for (a) pristine, (b) Li adsorbed, and (c) Cl adsorbed. The green, yellow, cyan, brown, pink spheres represent S, C, H, Li, and Cl atoms, respectively.}
\end{figure}

The structure under consideration for the monolayer polythiophene is depicted in Fig.\ 1(a). The supercell contains 16 carbon (C) atoms, 8 Hydrogen (H) atoms, and 4 Sulphur (S) atoms. Figs.\ 1(b) and (c) represent the Lithium (Li) and Chlorine (Cl) atoms adsorbed monolayer polythiophene. Structural parameters, such as bonding between C$-$C, C$-$H, C$-$S, and the bond angles are summarized in Table 1. Our calculated parameters agree well with previously reported experimental and theoretical values.\cite{Casado,jcp,Lischka} Small changes in the bond lengths and bond angles are found after allowing Li to adsorb on monolayer polythiophene. The Li atoms are located above the centre of the pentagon made from C and S atoms. The S-Li distance is 2.376 \AA, while the C-Li distances are found to be 2.171-2.235 \AA, see Table 1. The optimized structure of the Cl adsorption in monolayer polythiophene is shown in Fig.\ 1(c). A S-Cl distance of 3.675 \AA\ and C-Cl distances of 3.210-3.490 \AA\ are found; these values agree well with previous reports.\cite{jcp,Lischka} The Cl atom does not sit at the centre of the pentagon unlike Li adsorption. This can be understood from the fact that the Cl atom becomes negatively charged because of the charge transfer from polymer to Cl atom and is repelled by the slight negatively charged C atoms around it. The closest H-Cl distance is found to be about 2.56 to 2.77 \AA, which agrees well with previous findings. \cite{Zamoshchik1} 

The structures under consideration for bulk or bilayer polythiophene are addressed in Fig.\ 2. An interlayer spacing of 3.372 \AA\ or 3.371 \AA\ is found for bulk or bilayer polythiophene, which agrees well with other bilayer systems composed of C atoms. \cite{prl,prl-timm,cpl1,epl1,epl2,scirep,jmc1} The top and side views of the optimized structures are shown in Fig.\ 2(a). The other structural parameters are found to be similar to the monolayer case. The optimized structure of Li-intercalated bulk polythiophene is presented in Fig.\ 2(b). It is found that the interlayer spacing between two polythiophene layers varies from 3.653 \AA\ to 3.881 \AA. The spacing close to the area where Li-intercalated is found to be larger as compared to areas far from Li intercalation as expected. Unlike in the case of Li adsorbed monolayer polythiophene, the Li atoms are slightly de-attached from the S atom by forming C-S distances of 2.752 to 2.770 \AA, see Fig.\ 1(b) and 2(b) for comparison. The Li atom occupies a site exactly between the two polythiophene layers with a distance of 1.944 \AA\ from each of the two layers. The C$-$C bond lengths are found to be 1.392 to 1.421 \AA, a slight modification as compared to Li adsorbed monolayer polythiophene. Whereas, in case of Li-intercalated bilayer polythiophene, the separation between two polythiophene layer is found to be 3.763 \AA\ to 4.021 \AA\ with the distance of Li from each of two layers is found to be 2.012 \AA. The C$-$C bond length of 1.392 to 1.421 \AA\ with C$-$S bond length of 1.743 to 1.765 \AA\ are achieved, close to the bulk case. The obtained values of the interlayer spacing and the C$-$C bond lengths agree well with the experimentally observed values for Li-intercalated graphite.\cite{nature-physics,ACS-Nano,Thinius}. In addition, Li plays a crucial role to induce low temperature superconductivity in carbon-based materials. \cite{su11,su2,su3}

\begin{figure}[h]
\includegraphics[width=.95\textwidth,clip]{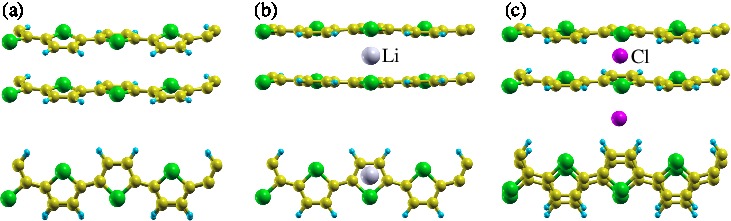}
\caption{Top and side views of the optimized structure of periodic bulk or bilayer polythiophene for (a) pristine, (b) Li intercalate, and (c) Cl intercalate, where bulk is defined as the periodic repetition along the $z$-direction.}
\end{figure}

The top and side views of the optimized structure of the Cl-intercalated bulk or bilayer polythiophene are depicted in Fig.\ 2(c). In case of bulk Cl-intercalated polythiophene, the Cl atoms do not sit between the two layers of the polythiophene but move away from the centre unlike for Li intercalation, which can again be attributed to the fact that the Cl atoms become negatively charged and are repelled by the slight negatively charged C atoms around them. The details of the charge analysis will be presented in the following section. Interestingly, the second layer of polythiophene shifts slightly as compared to the corresponding first layer, Fig.\ 2(c). After Cl intercalation the separation between two polythiophene layers is found to be 3.774 \AA\ to 3.812 \AA, larger than Li intercalation because the Cl atom has a larger ionic radius as compared to the  Li atom resulting in larger separation. The obtained values for the C$-$C bond length vary between 1.393 to 1.431 \AA. The C$-$S bond length is 1.744 \AA\ and the distance from the polythiophene layer to Cl is found to be 1.881 \AA, close to the Li case. In addition, the Cl-H distance is found to be 1.651 \AA, see the top view structure presented in Fig.\ 2(c), which indicates that Cl interacts with the H atoms. An interlayer spacing of 3.783 to 3.801 \AA\ is obtained for 
Cl-intercalated bilayer ploythiophene. C$-$C bond lengths of 1.402 to 1.423 \AA, a C$-$S bond length of 1.750 \AA, and a Cl-H distance of 1.931 \AA\ (larger as compared to bulk) are obtained. The distance between Cl and the adjacent polythiophene layer is found to be 1.931 \AA. In addition, the flipping of the second polythiophene layer by 180$^\circ$ has been considered in order to create bulk or bilayer structures. Comparing these to parallel bulk or bilayer, it is found that the parallel bulk or bilayer structures are energetically more favorable. The total ground state energy difference between parallel and flipped by 180$^\circ$ bulk or bilayer is found to be $-0.69$ eV and $-$1.12 eV, respectively, which clearly indicates that parallel bulk or bilayer polythiophene is energetically favorable. Similarly, the parallel bulk or bilayer polythiophene is stable by $-$1.09 eV or $-$0.88 eV and $-$1.03 eV or $-$0.36 eV after Li and Cl intercalation. Therefore, in the following, only the parallel bulk or bilayer is considered.  

\subsection{Electronic structure}
\begin{figure}[ht]
\includegraphics[width=0.7\textwidth,clip]{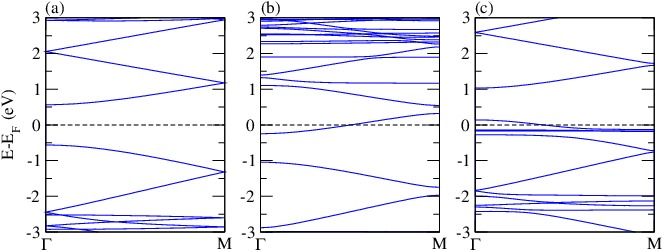}
\caption{The electronic band structure for monolayer of periodic (a) pristine polythiophene, (b) Li adsorbed polythiophene, and (c) Cl adsorbed polythiophene.}
\end{figure}

The electronic band structures for pristine, Li adsorbed, and Cl adsorbed monolayer polythiophene are addressed in Fig.\ 3. For pristine monolayer polythiophene, a band gap of 1.08 eV is obtained, falling exactly at the Fermi level (see Fig.\ 3(a)), which agrees well with previous findings.\cite{Asaduzzaman,jcp,Rittmeyer} Applying L\"owdin population analysis, it is found that the C atoms labelled as C5 and C2 absorb 0.09 electron, while S1 looses 0.41 electrons. Similarly, C4 and C3 become positively charged by an amount of 0.17 electron and the H atoms are negatively charged by 0.13 electron. This charge distribution indicates that there is some ionic character in the bonding throughout the network. In Fig.\ 3(b), the electronic band structure of Li adsorbed monolayer polythiophene is shown. It is found that the band gap of the pristine case shifts below the Fermi level by 0.68 eV by inducing $n$-doping.\cite{jcp} A back voltage could be used to shift the gap back to the Fermi level. The Li atom donates an electron to the pentagonal network of the polythiophene. As a result the Fermi level is raised and hence $n$-doping is found. The charge analysis shows that the Li atom loses 0.91 electron while S1 absorbs 0.17 electron, C5 and C2 absorb 0.15 electron each, and C3/C4 absorb 0.24/0.29 electron, which indicates that a significant charge redistribution has occurred as compared to pristine monolayer polythiophene. The $p$-doped states are observed when allowing a Cl atom to adsorb onto the monolayer polythiophene, see Fig.\ 3(c). The gap of pristine polythiophene shifts above the Fermi level by 0.61 eV upon Cl adsorption. The Cl atom gains a charge of 0.54 electron, C2/C5 lose a charge of 0.05 electrons, and C3/C4 lose about 0.11 electron, while the charges on the S and H atoms remain the same as in as pristine monolayer polythiophene. The binding energy was found to be $-2.01$ eV and $-1.49$ eV for Li and Cl adsorbed polythiophene, respectively, which agrees well with the Li adsorbed polythiophene in different concentrations.\cite{jcp} The binding energy is found to smaller by an amount of 0.101 eV and larger by 0.152 eV for 2 Li and 2 Cl adsorbed polythiophene, respectively, which indicates that the formation of a polaron in Li adsorbed polythiophene is more stable than that of a bipolaron while the situation is reverse for Cl adsorption. These findings are in line with previous reports for Li \cite {jcp} or Cl \cite{Zamoshchik1} adsorbed polythiophene. Moreover, for Li adsorption, a polaron and bipolaron induce gaps (below the Fermi level) of 0.82 eV and 0.73 eV, respectively. The obtained nature of the bands and magnitude of the gap agree well with a previous report.\cite{jcp} Whereas, for Cl adsorption, a polaron and bipolaron produce slightly larger gaps (above the Fermi level) of 0.93 eV and 0.88 eV, respectively, quantitative agreement is found with the replacement of C by Cl atom in polythiophene.\cite{Rittmeyer} In addition, experimentally, it has been demonstrated that the thermal properties (electrical conductivities) of the polythiophene can greatly be increased by Cl doping.\cite{Kelkar}

\begin{figure}[ht]
\includegraphics[width=0.8\textwidth,clip]{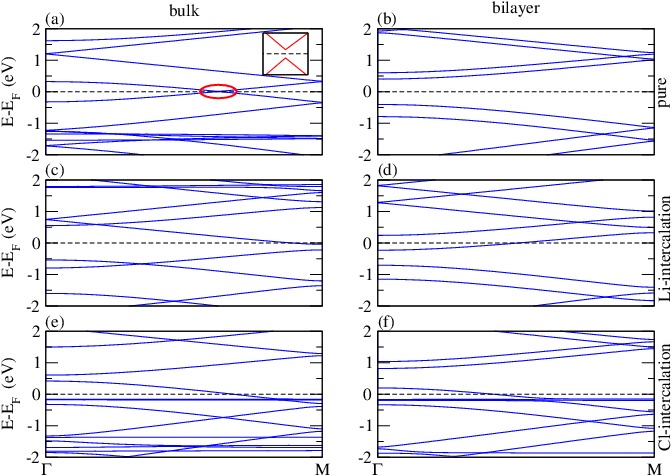}
\caption{The electronic band structure for periodic bulk and bilayer polythiophene: (a-b) pristine, (c-d) Li intercalation, and (e-f) Cl intercalation.}
\end{figure}
The electronic band structures of pristine bilayer, pure (pristine) bulk, Li/Cl-intercalated bulk, and Li/Cl-intercalated bilayer polythiophene are depicted in Fig.\ 4. For pristine polythiophene bulk, a very small band gap or about 4 meV (see inset in Fig.\ 4(a)) is obtained at the Fermi level, while a band gap of 0.81 eV is achieved for the bilayer, which is smaller than that of the monolayer. There are two well separated bands close to the conduction regime and two bands close to the valence regime above/below the Fermi level, see Fig.\ 4(b). The results obtained from this investigation are consistent with previous experimental and theoretical reports, the band gap can be narrowed down either by increasing the length of the polymer chain or by $\pi$-stacking of the polymer. \cite{Houk,Nakano,Yu1} Both techniques result in the evolution of extra bands at the Fermi level, thus, reduction in the band gap can be obtained. The gap between the two bands is about 0.41 eV, and due to the splitting of the bands the band gap in the bilayer reduces as compared to the corresponding monolayer case, as expected. The charge distribution in this case is the same as for the monolayer. After Li intercalation, the system becomes $n$-doped by shifting the gap of the pristine case below the Fermi level by 0.25 eV with reduction in the gap of about 0.24 eV, see Fig.\ 4(c). Similarly, shifting of the gap by 0.52 eV with a gap of 0.51 eV is obtained for Li-intercalated bilayer polythiophene, see Fig.\ 4(d). At this point, it is important to mention that there are two possible optical transitions during light absorption and exciton formation in the system; they can be described as $\pi$-$\pi$ transitions. Experimentally, such a behavior has been detected for doped $\pi$-conjugated polymers.\cite{Gasiorowski} In case of a doped polymer, where the free charges are injected onto the polymer, polaronic features are obtained. The doping can be understood by two transitions with energies in the range of IR polaronic absorption (small gap) and optical polaronic absorption (larger gap). Normally, the maximum in-gap transitions have been found to be 0.45-0.85 eV and 1.3-1.7 eV \cite{Gasiorowski}, respectively. The values obtained in this study amount to 0.45 eV and 0.96 eV for IR polaronic absorption and optical polaronic absorption, respectively. Quantitatively, this behavior is consistent for all the systems under study. Moreover, after doping, the conductivity can be increased strongly and hence, the systems would be potentially useful for devices with high conductivity, such as photovoltaics, thermoelectrics, and spintronics.\cite{Gasiorowski,Gagnon1,Braun}

In both cases, the Li atom loses about 0.84 electron, while the S atom gains 0.05 electron, C2/C5 gain 0.21 electrons, whereas the charges on the other atoms remain the same as for their pristine counterparts. For Cl-intercalated bulk or bilayer polythiophene (Figs.\ 4e and f), a shift of the gap as compared the pristine case by about 0.51 eV or 0.50 eV above the Fermi level is found, which makes the system $p$-doped. A gap of 0.21 eV or 0.62 eV is achieved for the bulk or bilayer case, respectively. Based on L\"owdin population analysis, the Cl atom gains a charge of about 0.64/0.67 electron for the bulk/bilayer case, while the H atoms gain a small charge of about 0.02 electron and C2/C5 lose charge of about 0.03 electron as compared to the pristine counterpart. Moreover, the calculated value of the binding energy is found to be $-$2.29 eV or $-$2.77 eV for Li-intercalated bulk or bilayer polythiophene, which indicates that the Li intercalation in the bilayer case is energetically favorable by 0.48 eV. Whereas, the binding energy is obtained to be $-$1.62 eV or $-$1.84 eV for Cl-intercalated bulk or bilayer polythiophene; again the data indicates that Cl intercalation in the bilayer case is energetically favorable by 0.21 eV. Moreover, the obtained value of the binding energy in case of bulk or bilayer is higher than that of monolayer polythiophene, which reflects that the binding of Li or Cl in bulk or bilayer is strong than that of monolayer polythiophene. In addition, the effect of an external electric field along the $x$- as well as the $z$-direction has been examined for both monolayers and multilayers (for both the doped and undoped cases) according to the scheme described in Refs.\ \cite{e1,e2,e12} The magnitude of the band gap is found to be exactly the same for both directions of the electric field. The magnitude of the applied external electric field is $1\cdot 10^8$ V/m, which is easily accessible in the experiment. The band gap is increased only by only $\sim 1.7$ \% in all the cases under study.

\section{Molecular calculations}
\subsection{Monolayer polythiophene}
\begin{figure}[ht]
\includegraphics[width=0.8\textwidth,clip]{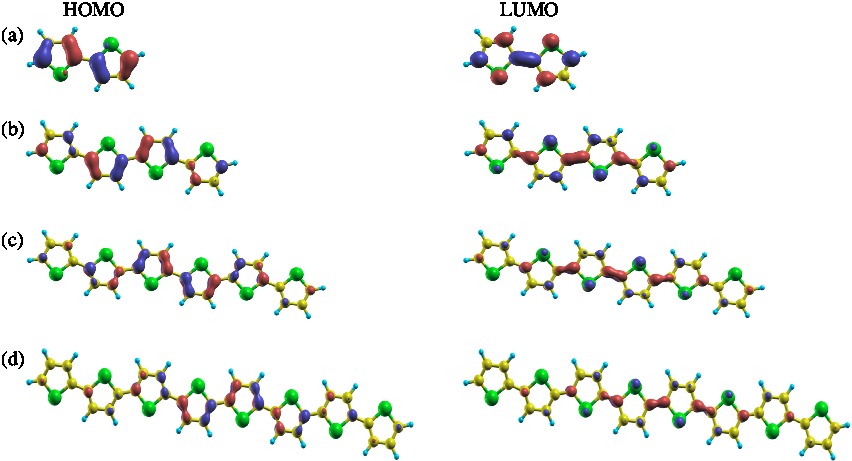}
\caption{The calculated HOMO and LUMO with an isovalue of $\pm 0.05$ electrons/\AA$^3$ of polythiophene molecule for (a) 2-ring, (b) 4-ring, (c) 6-ring, and (d) 8-ring. The green, yellow, and cyan spheres represent S, C, and H atoms, respectively.}
\end{figure}

\begin{table}[ht]
\begin{tabular}{|c|c|c|c|c|c|c|c|c|c|c|}
\hline
molecule&HOMO     &LUMO    &$E_{gap}$ &C$-$C     &C$-$S     &C$-$H &C-S-C &C-C-C   &S-C-C \\
\hline
2-ring&$-5.01$  &$-2.07$ &2.94  &1.37-1.45 &1.70-1.72 &1.08  &93    &111-113 &110    \\
\hline
4-ring &$-4.60$  &$-2.73$ &1.87  &1.38-1.45 &1.36-1.37 &1.08  &110   &107-108 &107     \\
\hline
6-ring&$-4.48$  &$-2.95$ &1.54  &1.38-1.44 &1.38-1.39 &1.08  &111   &107-108 &106      \\
\hline
8-ring&$-4.44$  &$-3.05$ &1.39  &1.38-1.43 &1.37-1.38 &1.08  &111   &108     &106       \\
\hline
\end{tabular}
\caption{The calculated energy of HOMO and LUMO (in eV), HUMO-LUMO gap (in eV), bond lengths (in \AA), and bond angles (in $^\circ$).} 
\end{table} 
For the molecular calculations, four different structures are considered as depicted in Fig.\ 5 with their corresponding HOMO and LUMO orbitals, namely 2-ring, 4-ring, 6-ring, and 8-ring. The HOMO and LUMO energies, HOMO-LUMO gap, bond distances, and bond angles for all the systems are listed in Table 2. The 2-ring structure is a rather small model for polythiophene, as indicated by the larger bond length and larger HOMO-LUMO gap. Upon increasing the number of rings to 4, 6, and 8, the structural parameters and band gap become smaller and show convergence. This fact has already been addressed in many theoretical predictions and experimental realizations.\cite{Kasai,Bendikov,Camarada,Mannsfeld,Autschbach}  Control of the band gap in $\pi$-conjugated polymers such as polythiophenes is important because low band gap conjugated polymers are of great interest for their intrinsic conductivity in electronic devices, in particular, light emitting diodes \cite{Reecht} and solar cells. \cite{woo} The HOMO-LUMO gap is found to be 2.94 eV for the 2-ring system and decreases to 1.87 eV to 1.54 eV to 1.39 eV for 4-ring to 6-ring to 8-ring, respectively. Essentially, this demonstrates that the 6-ring and 8-ring systems are similar in nature from an electronic point of view. The HOMO-LUMO gap can can be modified either by creating disorder \cite{Vukmirovi} in the polythiophene or by doping \cite{Wudl,Dinga,Takechi}, such that the modified polythiophene can be used in electronic devices. Quantitatively, the calculated value of the band gap in the periodic case is similar to the HOMO-LUMO gap of the 6-ring or 8-ring.

\begin{figure}[ht]
\includegraphics[width=0.5\textwidth,clip]{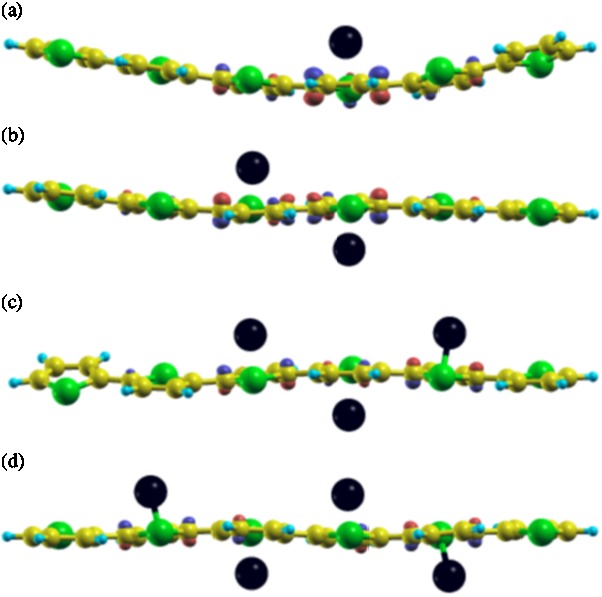}
\caption{Optimized structures of 6-ring polythiophene with coverage of (a) 1 Li atom, (b) 2 Li atoms, (c) 3 Li atoms, and (d) 4 Li atoms with corresponding HOMO.}
\end{figure}

\begin{figure}[ht]
\includegraphics[width=0.5\textwidth,clip]{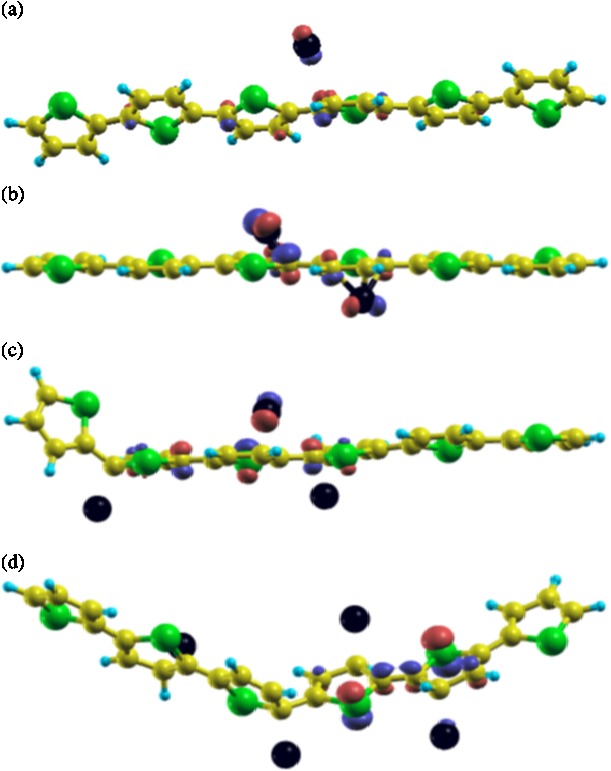}
\caption{Optimized structure of 6-ring polythiophene with coverage of (a) 1 Cl atom, (b) 2 Cl atoms, (c) 3 Cl atoms, and (d) 4 Cl atoms together with calculated HOMO. }
\end{figure}  
In the following, the 6-ring structure of polythiophene is used in order to adsorb Li or Cl atoms. The side view of the optimized structures under consideration with corresponding HOMOs are shown in Figs.\ 6(a-d) and Figs.\ 7(a-d) for Li or Cl adsorption, respectively. Essentially, minor modifications in the structural parameters such as bond lengths and bond angles, as well as corrugation in the structure, are found as compared to the un-adsorbed 6-ring of polythiophene. The Li atom is found to sit at a site at the centre of the pentagon similar to the case Li adsorbed periodic polythiophene, see Fig.\ 6(a). Based on Mulliken charge analysis, the Li atom transfers a charge of 0.502 electron to the pentagonal ring of the polythiophene. The calculated binding energy of a Li atom onto the polythiophene molecule is found to be $-$28.6 Kcal/mol with a HOMO-LUMO gap of 0.775 eV. In the case of adsorption of 2 Li atoms, two different structures are considered, such that one Li atom lies at positive \textit{z}-direction and the other lies at negative \textit{z}-direction (above and below the molecular plane, updn) or both Li atoms lie on the same side (upup or dndn) with respect to the polythiophene. It is found that the former case is energetically favourable by 0.838 eV. Thus, the former configuration is used for further study; its optimized geometry is shown in Fig.\ 6(b). A binding energy of $-$28.6 Kcal/mol per atom, an average amount of the charge transfer of 0.497 electron, and a HOMO-LUMO gap of 0.772 eV are obtained. Similarly, the updnup configuration is energetically more favorable by 0.752 eV as compared to the upupup configuration for 3 Li atoms adsorbed on the 6-ring structure of polythiphene. Strong modification in the structure is obtained, see Fig.\ 6(c). A binding energy per atom of $-$28.0 Kcal/mol is found, which is slightly smaller than that of 2 Li atom adsorption. The average amount of charge transfer from the Li atoms to the molecule is found to be 0.487 electron with a HOMO-LUMO gap of 0.653 eV. Similarly, the updnupdn configuration is energetically favorable over upupupup by an energy of 0.784 eV; the side view of the optimized structure is shown in Fig.\ 6(d). A binding energy of $-$28.0 Kcal/mol is found, which is the same as in the case of 3 Li atoms adsorption. In this case, an average charge transfer of 0.478 electron from the Li atoms is obtained. The HOMO-LUMO gap is 0.658 eV, which indicates that the HOMO-LUMO gap is decreasing with increasing the adsorbent coverage. Moreover, the polythiophene with a single polaron is found to be more stable then the bipolaron, tripolaron, and tetrapolaron, which follows from the binding energies presented above. This observation agrees well with a previous report.\cite{jcp} However, the energy difference between single polaron and multipolaron is very small. The small energy difference indicates that clustering of the dopant can be considered at finite temperature. For conducting polymers, it has been proposed that large increments in the conductivity can be achieved in the presence of polarons and bipolarons.\cite{street,Pomerantz}

Strong modifications in the structural parameters are observed for Cl atom adsorption onto the 6-ring polythiophene. Similar to the case of Cl adsorbtion in the periodic calculations, the Cl atom is found to move away from the polythiophene molecule but still interact with H atoms. The side view of the optimized structure is depicted in Fig.\ 7(a). The Mulliken charge analysis shows that the Cl atom gains a charge of about 0.452 electron from the polythiophene molecule. The binding energy of 1 Cl atom onto the polythiophene molecule is $-$26.4 Kcal/mol, quite close to the case of Li adsorption. The HOMO-LUMO gap is found to be 1.740 eV, twice than that of polythiophene with 1 Li atom adsorbed. In this case the polythiphene molecule is buckled, which essentially provides an internal electric field. The electric field is responsible for opening the band gap (HOMO-LUMO gap), which has already been demonstrated for carbon and silicon based systems.\cite{Ashwin,georg} For 2 Cl atoms, the updn configuration is energetically favorable by an energy of 0.811 eV. The side view of the optimized structure is shown in Fig.\ 7(b). A binding energy of $-$28.6 Kcal/mol is obtained. The average amount of the charge gained by the Cl atoms is found to be 0.431 electron, and the HOMO-LUMO gap is 1.327 eV. For 3 Cl adsorption, the updnup configuration is energetically favourable with an energy difference of 0.701 eV relative to the upupup configuration. One of the pentagonal rings is corrugated and moves up to the $xy$-plane, see Fig.\ 6(c). The obtained value of the binding energy amounts to $-$29.0 Kcal/mol, slightly higher than that of 2 Cl atom adsorption. The average amount of charge gained by the Cl atoms is 0.432 electron, and the HOMO-LUMO gap is 1.078 eV. Finally, for 4 Cl atom adsorption, the updnupdn configuration is energetically favorable, as in the case of Li adsorption, by an energy of about 0.799 eV, see the side view of the optimized structure in Fig.\ 7(d). A binding energy of $-$33.0 Kcal/mol is found, which is higher than that of the 3 Cl atoms adsorption. An average charge gain of about 0.421 electron and a HOMO-LUMO gap of 1.667 eV are found.

In addition, for comparison, the calculations for (6-ring)$^-$ and (6-ring)$^+$ are performed. In principle, these are similar to calculations for Li and Cl doping in 6-ring polythiophene, at least in as far as the interactions between the polythiophene and the dopant are ionic in nature. The obtained values of the HOMO-LUMO gap are found to be 0.523 eV and 1.309 eV, respectively, close to the Li or Cl doping in 6-ring polythiophene as expected. In contrast to the Li adsorbed polythiophene, our results prevails that a single polaron is less stable than that of bipoloran, which is in the line of chlorination of polythiophene. \cite{Zamoshchik1} Experimentally and theoretically, in the case of Cl doping, it has been observed that bipolaron is energetically more preferable for high dopant concentrations i. e. one dopant per six or less thiophene rings, while a single polaron is energetically more favorable for low dopant concentrations i. e. one dopant per ten or more thiophene rings. \cite{Cooney,Zamoshchik1}

\subsection{Bilayer polythiophene}

\begin{figure}[ht]
\includegraphics[width=0.8\textwidth,clip]{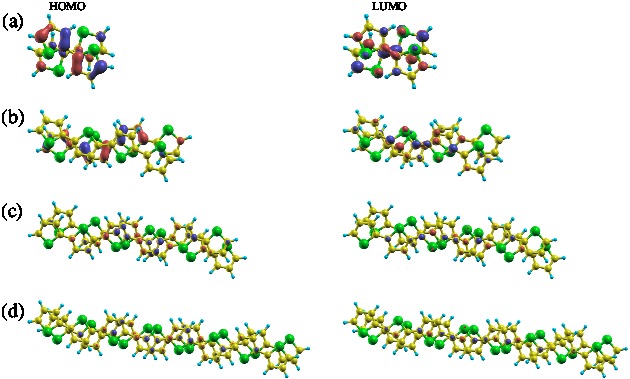}
\caption{The calculated HOMO and LUMO with an isovalue of $\pm 0.05$ electrons/\AA$^3$ of bilayer polythiophene molecule for (a) 2-ring, (b) 4-ring, (c) 6-ring, and (d) 8-ring. The green, yellow, and cyan spheres represent S, C, and H atoms, respectively.}
\end{figure}

The focus of this section are the bilayer of polythiophene in a parallel configuration and with the second layer flipped by $180^\circ$. Experimentally, it has been demonstrated that the stacking of the $\pi$ conjugated thiophene is useful in order to obtain long-range charge and energy transport in novel nanoscale polymer based devices.\cite{grey2} The optimized structures of parallel bilayer polythiophene are addressed in Fig.\ 8 with the corresponding HOMO and LUMO. The parallel bilayer structures are energetically favorable by small amounts of energy of about 4.31 meV, 8.63 meV, 63.34 meV, and 115.84 meV for 2-ring, 4-ring, 6-ring, and 8-ring, respectively. The small energy differences between these two types of bilayers indicate that the flipped bilayer can also be stable, thus, it should be feasible to synthesize such structures. It is found that the second layer is slightly shifted with respect to the first layer in both the parallel bilayers as well as in the flipped bilayers of all sizes, which agrees well with the experimental realization of the $\pi$ stacked polymers. \cite{yu2} Apart from that minor modifications in the structural parameters are found as well. Selected structural parameters are summarized in Table 3. The obtained value of the interlayer distance agrees well with the experimentally and theoretically observed values for stacked molecules of similar geometry.\cite{grey3} The obtained value for the interlayer distance also agrees well with those of carbon based multilayers.\cite{Bokdam,jmc2,Brihuega} The HOMO-LUMO gap decreases in both types of bilayers as compared to that of the monolayer, which can be attributed to the interlayer interactions and the resulting band splitting at the Fermi level. The HOMO-LUMO gap varies from 1.989 eV to 1.200 eV and 2.719 eV to 1.338 eV for 2-ring to 8-ring in case of the parallel bilayers and flipped bilayers, respectively, see Table 2. It is worth to mention that the HOMO-LUMO gap in case of the flipped bilayers is larger than that of the parallel bilayers, which can be attributed to weaker interlayer interactions and hence weaker band splitting. Experimentally, it has been proposed that such stacked systems are of potential use for optical or electronic switches.\cite{grey4}

\begin{table}[h]
\begin{tabular}{|c|c|c|c|c|c|c|c|c|}
\hline
System& \multicolumn{4}{|c|}{\multirow{1}{*}{Parallel bilayer}}&\multicolumn{4}{|c|}{\multirow{1}{*}{Flipped bilayer}}\\
\cline{2-7}
\hline
&HOMO (eV)&LUMO (eV)&E$_{gap}$ (eV)&$d_{av}$ (\AA)&HOMO (eV)&LUMO (eV)&E$_{gap}$ (eV)&$d_{av}$ (\AA) \\
\hline
2-ring&$-4.731$&$-2.747$&1.989&3.45&$-4.687$&$-1.968$&1.468&3.48 \\
\hline
4-ring&$-4.371$&$-2.557$&1.814&3.58&$-4.441$&$-2.540$&1.901&3.62 \\
\hline
6-ring&$-4.127$&$-2.768$&1.359&3.55&$-4.243$&$-2.775$&1.719&3.56 \\
\hline
8-ring&$-4.071$&$-2.871$&1.200&3.57&$-4.224$&$-2.886$&1.338&3.59 \\
\hline
\end{tabular}
\caption{The calculated HOMO and LUMO energies, the HOMO-LUMO gap, and the average interlayer distance for parallel bilayers and flipped bilayers.} 
\end{table}

\begin{figure}[ht]
\includegraphics[width=0.5\textwidth,clip]{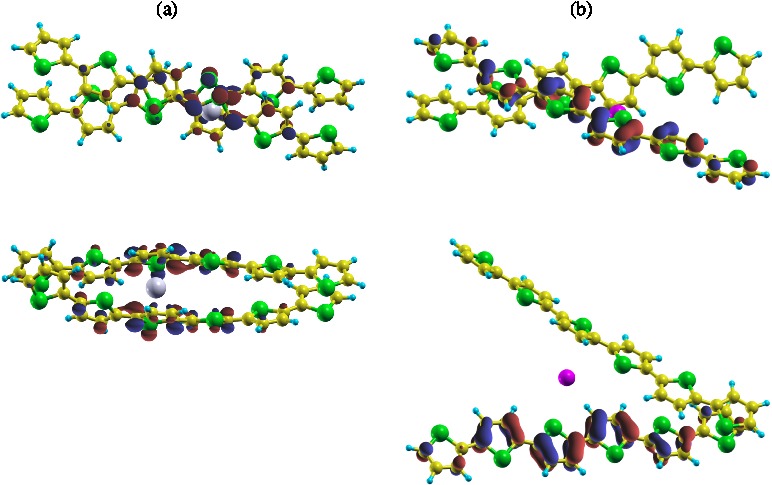}
\caption{The optimized structures (top and side views) of (a) Li-intercalated polythiophene and (b) Cl-intercalated polythiophene with corresponding HOMO (isovalue of $\pm 0.05$ electrons/\AA$^3$) for the 6-ring bilayer.}
\end{figure}

For the Li or Cl intercalation between the two layers, a 6-ring polythiophene is used, see the top and side views shown in Fig.\ 9(a-b). The Li or Cl-intercalated in the parallel bilayer of polythiophene is energetically more favourable compared to the flipped bilayer by 9.21 meV or 12.34 meV. It is found that the Li atom occupies a site in the middle of the two polythiophene layers and both the layers are shifted strongly with respect to each other. The HOMO is found to be located close to the Li atom but delocalized over both layers. 
A charge transfer of 0.391 electron is found to both the polythiophene layers. The HOMO-LUMO gap is 0.275 eV. In case of Cl atom intercalation between the polythiophene molecules, a strong corrugation as well as shifting of the two layers with respect to each other occurs, which significantly modifies the structural parameters and interlayer spacing between the layers, see top and side views presented in Fig.\ 9(b). The energy of switching the undoped to the doped state amounts to 1.17 eV, which is smaller than the energy associated (2.28 eV) with the case of Li intercalation between the two polythiophene layers. Moreover, experimentally, Cl doped graphite has been studied by scanning tunneling microscopy, Raman spectroscopy, and Hall measurements. Strong distortions in the graphene layers of the graphite were found \cite{carbon}, which is in line with our observations for Cl intercalated polythiophene. Unlike for the Li case, the HOMO is found to be located only on the bottom layer of polythiophene. In this case, the Cl atom gains a charge of 0.572 electron and the HOMO-LUMO gap is 0.031 eV. 

\section{Conclusion}
Using first-principles density functional theory based calculations, the structural and electronic properties of polythiophene and its derivatives in periodic and oligomer forms have been investigated. The effect of Li or Cl adsorption onto a monolayer and Li or Cl-intercalated into bulk or bilayer polythiophene in the periodic form have been studied. These systems are extremely useful to construct polythiophene based electronic and optoelectronic devices. The calculated binding energy for Li or Cl adsorbed bulk or bilayer polythiophene is found to be larger than that of the corresponding monolayers. Furthermore, the oligomeric forms of all the periodic systems were also studied. The trends in the binding energy as a function of the adsorbent remain the same in the case of the molecular forms. The calculated band gap (in periodic calculations), the HOMO-LUMO gap (in the case of the molecular calculations), as well as the charge transfer are analyzed in detail. Moreover, in cases of bulk or bilayer, different kinds of stacking have been considered. The parallel bulk or bilayer structures are energetically favorable as compared to flipping the second layer by 180$^\circ$ (in order to create the bulk or bilayer). However, the latter structures are still energetically accessible at ambient temperature. This type of configuration has been considered for both the periodic as well as oligomer forms. The obtained results for the monolayer systems with and without Li or Cl adsorption agree well with previous reports. Moreover, the obtained data revealed that in case of Li adsorbed polythiophene, polarons are more stable than bipolarons, while for Cl adsorption bipolarons are energetically more favorable, in good agreement with a previous report.\cite{jcp,Zamoshchik1} The formation of the polarons or multipolarons is very important in order to achieve larger conductivity in Li or Cl adsorbed polythiophene.\cite{street,Pomerantz} It is believed that the detailed analysis of the present investigations would be useful to understand the structural properties and the tuneability of the electronic states, which could be an important step toward constructing polythiophene based electronic devices. 

\section{Acknowledgement}
GS acknowledges funding from the Natural Sciences and Engineering Council of Canada (NSERC, Discovery Grant). MSF acknowledges support by the Natural Sciences and Engineering Research Council (NSERC) of Canada, the Canada Research Chair program, Canada Foundation for Innovation (CFI), the Manitoba Research and Innovation Fund, and the University of Manitoba.

\end{document}